\documentclass{lipics-v2021}

\usepackage{graphicx}
\usepackage{color}
\usepackage{enumerate}
\usepackage[linesnumbered]{algorithm2e}
\usepackage{amsmath}
\usepackage{amssymb}
\usepackage{tikz}
\usetikzlibrary{calc}
\newcommand{\hcancel}[5]{%
    \tikz[baseline=(tocancel.base)]{
        \node[inner sep=0pt,outer sep=0pt] (tocancel) {#1};
        \draw[red] ($(tocancel.south west)+(#2,#3)$) -- ($(tocancel.north east)+(#4,#5)$);
    }%
}%

\newcommand{\dd}{\mathinner{.\,.}}

\newcommand{\shorten}[1]{}
\newcommand{\no}[1]{}

\title{BAT-LZ Out of Hell} 

\author{Zsuzsanna Lipt\'ak}{Dipartimento di Informatica, University of Verona,
Italy}{zsuzsanna.liptak@univr.it}{https://orcid.org/0000-0002-3233-0691}{Partially funded by the MUR PRIN project Nr.\ 2022YRB97K 'PINC' (Pangenome INformatiCs. From Theory to Applications) and
by the INdAM-GNCS Project CUP$\_$E53C23001670001 (Compressione, indicizzazione, analisi e confronto di dati biologici).}
\author{Francesco Masillo}{Dipartimento di Informatica, University of Verona,
Italy}{francesco.masillo@univr.it}{https://orcid.org/0000-0002-2078-6835}{}
\author{Gonzalo Navarro}{Center for Biotechnology and Bioengineering (CeBiB) \\
Department of Computer Science, University of Chile,
Chile}{gnavarro@dcc.uchile.cl}{https://orcid.org/0000-0002-2286-741X}{Funded by Basal Funds FB0001, Mideplan,
Chile, and Fondecyt Grant 1-230755, Chile.}

\authorrunning{Zs.\ Lipt\'ak, F.\ Masillo, G.\ Navarro} 
\Copyright{Zsuzsanna Lipt\'ak, Francesco Masillo, Gonzalo Navarro}

\category{} 

\relatedversion{} 

\supplement{}
\supplementdetails[subcategory={Source Code}, cite={}, swhid={}]{Software}{https://github.com/fmasillo/BAT-LZ}

\nolinenumbers 

\EventEditors{Shunsuke Inenaga and Simon J. Puglisi}
\EventNoEds{2}
\EventLongTitle{35th Annual Symposium on Combinatorial Pattern Matching (CPM 2024)}
\EventShortTitle{CPM 2024}
\EventAcronym{CPM}
\EventYear{2024}
\EventDate{June 25--27, 2024}
\EventLocation{Fukuoka, Japan}
\EventLogo{}
\SeriesVolume{296}
\ArticleNo{17}

\newcommand{\SA}{\mathsf{SA}}
\newcommand{\ISA}{\mathsf{ISA}}
\newcommand{\ST}{\mathsf{ST}}

\newcommand{\minmax}{\mathsf{minmax}}
\newcommand{\txtpos}{\mathsf{txtpos}}
\newcommand{\real}{\mathsf{real}}

\newcommand{\leaf}{\ensuremath \textit{leaf}}
\newcommand{\parent}{\ensuremath \textit{parent}}
\newcommand{\sd}{\textit{sd}}

\newcommand{\BLZ}{\text{BAT-LZ}}

\def\a{{\tt a}}
\def\b{{\tt b}}
\def\d{{\tt d}}
\def\l{{\tt l}}
\def\r{{\tt r}}

\begin{document}

\maketitle 

\bigskip

\begin{abstract}
Despite consistently yielding the best compression on repetitive text collections, the Lempel-Ziv parsing has resisted all attempts at offering relevant guarantees on the cost to access an arbitrary symbol. This makes it less attractive for use  on compressed self-indexes and other compressed data structures. In this paper we introduce a variant we call BAT-LZ (for Bounded Access Time Lempel-Ziv) where the access cost is bounded by a parameter given at compression time. We design and implement a linear-space algorithm that, in time $O(n\log^3 n)$, obtains a BAT-LZ parse of a text of length $n$ by greedily maximizing each next phrase length. The algorithm builds on a new linear-space data structure that solves 5-sided 
orthogonal range queries in rank space, allowing updates to the coordinate where the one-sided queries are supported, in $O(\log^3 n)$ time for both queries and updates. This time can be reduced to $O(\log^2 n)$ if $O(n\log n)$ space is used. 

We design a second algorithm that chooses the sources for the phrases in a clever way, using an enhanced suffix tree, albeit no longer guaranteeing longest possible phrases. This algorithm is much slower in theory, but in practice it is comparable to the greedy parser, while achieving significantly superior compression. 
We then combine the two algorithms, resulting in a parser that always chooses the longest possible phrases, and the best sources for those. Our experimentation shows that, on most repetitive texts, our algorithms reach an access cost close to $\log_2 n$ on texts of length $n$, while incurring almost no loss in the compression ratio when compared with classical LZ-compression. Several open challenges are discussed at the end of the paper. 
 
\end{abstract}

\newpage

\setcounter{page}{1}

\section{Introduction}

The sharply growing sizes of text collections, particularly repetitive ones,
has raised the interest in compressed data structures that can maintain the
texts all the time in compressed form~\cite{Navacmcs20.2,Navacmcs20.3,Nav16}. For archival purposes, the original 
Lempel-Ziv (LZ) 
\shorten{\footnote{\shorten{Also referred to as LZ76, to distinguish it from LZ78, the dictionary-based compression algorithm by the same authors~\cite{ZL78}. 
It should also be noted that LZ76 is often misnamed LZ77, in reference to a 1977 paper~\cite{ZL77}, in which Lempel and Ziv presented a windowed version of the original algorithm~\cite{LZ76}.}}} 
compression format~\cite{LZ76} is preferred because it yields 
the least space among the methods that support compression and decompression 
in polynomial time---actually, Lempel-Ziv compresses and decompresses a text 
$T[1\dd n]$ in $O(n)$ time~\cite{RPE81}.
For using a compression format as a compressed data structure, however---in particular, to build a compressed text self-index on it~\cite{KN13}---, we 
need 
that arbitrary text snippets $T[i\dd i+\ell]$ can be extracted
efficiently, without the need of decompressing the whole text up to the desired
snippet. Grammar compression formats~\cite{KY00} allow extracting such text
snippets in time $O(\ell+\log n)$~\cite{BLRSRW15,GJL21}, which is nearly optimal
\cite{VY13}. So, although the compression they achieve is always lower-bounded 
by the size of the LZ parse~\cite{Ryt03,CLLPPSS05}, grammar compression 
algorithms are preferred over LZ compression in the design of text indexes 
\cite{Navacmcs20.3,CNPjcss21}, and of compressed data structures in general.

The LZ compression algorithm parses the text $T$ into a sequence of so-called
phrases, where each phrase points backwards to a previous occurrence of it in
$T$ and stores the next symbol in explicit form. While this yields a simple 
linear-time left-to-right decompression algorithm, consider the problem of 
accessing a particular symbol $T[i]$. Unless it is the final explicit symbol
of a phrase, we must determine the text position $j < i$ where $T[i] = T[j]$
was copied from. We must then determine $T[j]$, which again may be---with low
chance---the end of a phrase, or it may---most likely---refer to an earlier symbol
$T[j]=T[k]$, with $k<j$. The process continues until we hit an explicit
symbol. The cost of extracting $T[i]$ is then proportional to the length of
that {\em referencing chain} $i \rightarrow j \rightarrow k \rightarrow
\ldots$ Despite considerable interest in algorithms to access arbitrary text
positions from the LZ compression format, and apart from some remarkable
results on restricted versions of LZ~\cite{KS22}, there has been no progress
on the original LZ parse (which yields the strongest compression).

In this paper we introduce and study an LZ variant we call {\em Bounded Access Time 
Lempel-Ziv (BAT-LZ)}, which takes a compression parameter $c$ and produces a parse where no symbol has a
referencing chain longer than $c$, thereby guaranteeing $O(c)$ access time.\footnote{A parsing like \BLZ\ was described as a baseline in the experimental results in previous work~\cite{KNdcc10.1} of one of the authors, but without a parsing algorithm, see Sec.~\ref{sec:lz} for more details.} As opposed to classical LZ, BAT-LZ parses allow very fast access to the text, indeed, like a bat out of hell.

We design a {\em Greedy \BLZ\ parser}, which at each step of the
compression chooses the longest possible phrase. Finding such a phrase boils 
down to solving a 4-sided orthogonal range query in a 3-dimensional grid (in 
rank space), where one of the
coordinates undergoes updates as the parsing proceeds. We design such a data
structure, which turns out to handle 5-sided queries and support updates on
the coordinate where the query is one-sided. 
Our data structure handles queries and updates in time $O(\log^3 n)$, yielding a greedy \BLZ\ parsing 
in time $O(n\log^3 n)$ and space $O(n)$. 
We then design another \BLZ\ parser, referred to as {\em Minmax}, which runs on an enhanced suffix tree. It looks for the ``best'' possible sources of the chosen phrases, that is, with symbols having shorter referencing chains, while not necessarily choosing the longest possible phrase. Finally, we combine the two ideas, resulting in our {\em Greedier parser}, which runs again on an enhanced suffix tree. 
These last two algorithms, while their running time is upper bounded by $O(n^3\log n)$, both run in decent time in practice.

We implemented and tested our three \BLZ\ parsers on various repetitive 
texts of different sorts, comparing them with the original LZ parse and with 
two simple baselines that ensure \BLZ\ parses without any optimization.
The results show that all three algorithms run in a few seconds per megabyte and produce much better parses than the baselines. For values of $c=O(\log n)$ with a small constant, they produce just a small fraction of extra phrases on top of LZ. In particular, Greedier increases the size of the LZ parse by less than 1\% with $c$ values that are about $\log_2 n$ (i.e., 
20--30 in our texts). 

We note that, unlike the original LZ parse, a greedy parsing does not guarantee obtaining the minimal \BLZ\ parse. 
Indeed, finding the optimal \BLZ\ parse has recently been shown to be NP-hard for all constant $c$, and also hard to approximate for any constant approximation ratio~\cite{CicaleseU24}. Our results show that, on repetitive texts, a polylog-linear time greedy algorithm can nonetheless achieve good compression while guaranteeing fast access to text snippets. The other two algorithms are still polynomial time and offer fast access with almost no loss in compression compared to the classical LZ-compression. In our scenarios of interest (i.e., accessing the compressed text at random) the data is compressed only once and accessed many times, so slower compression algorithms can be afforded in exchange for faster access. 
We discuss at the end this and some other problems our work opens.


\section{Basic Data Structures}\label{sec:basics}

A string (or text) $T$ is a finite sequence of characters from an alphabet $\Sigma$. We write $T=T[1..n]$ for a string $T$ of length $n$, and assume that the final character is a unique end-of-string marker 
$\$$. We index strings from $1$ and write $T[i..j]$ for the substring $T[i]\dd T[j]$, $T[i\dd]$ for the suffix starting in position $i$, and $T[\dd j]$ for the prefix ending in position $i$.

\subparagraph*{Bitvectors and Wavelet Matrices.}
A bitvector $B[1\dd n]$ can be stored using $n$ bits, or actually $\lceil n/w \rceil$ words on a $w$-bit word machine, while providing access and updates to arbitrary bits in constant time. If the bitvector is static (i.e., does not undergo updates) then it can be preprocessed to answer $rank$ queries in $O(1)$ time using $o(n)$ further bits~\cite{Cla96,Mun96}:  $rank_b(B,i)$, where $b \in \{0,1\}$ and $0 \le i \le n$, is the number of times bit $b$ occurs in $B[1\dd i]$.

A wavelet matrix~\cite{CNO15} is a data structure that can be used, in particular, to represent a discrete $[1,n] \times [1,n]$ grid, with exactly one point per column, using $n\log_2 n + o(n\log_2 n)$ bits. Let $S[1\dd n]$ be such that $S[i]$ is the row of the point at column $i$. The first wavelet matrix level contains a bitvector $B_1[1\dd n]$ with the highest (i.e., $\lceil \log_2 n \rceil$th) bit of every value in $S$. For the second level, the sequence values are stably sorted by their highest bit, and the wavelet matrix stores a bitvector $B_2[1\dd n]$ with the second highest bits in that order. To build the third level, the values are stably sorted by their second highest bit, and so on. Every level $i$ also stores the number $z_i = rank_0(B_i,n)$ of zeros in its bitvector.

The value $S[i]$ can be retrieved from the wavelet matrix in $O(\log n)$ time. Its highest bit is $b_1 = B_1[i_1]$, with $i_1=i$. The second highest bit is $b_2 = B_2[i_2]$, with $i_2 = rank_0(B_1,i_1)$ if $b_1=0$ and $i_2 = z_1 + rank_1(B_1,i_1)$ if $b_1=1$. The other bits are obtained analogously.

The wavelet matrix can also obtain the grid points that fall within a rectangle $[x_1,x_2] \times [y_1,y_2]$ (i.e., the values $(i,S[i])$ such that $x_1 \le i \le x_2$ and $y_1 \le S[i] \le y_2$) in time $O(\log n)$, plus $O(\log n)$ per point reported. We start at the first level, in the range $B_1[sp_1,ep_1] = B_1[x_1,x_2]$. We then map the range into two ranges of the second level: the positions $i$ where $B_1[i]=0$ are all mapped to the range $B_2[sp_2,ep_2] = B_2[rank_0(B_1,sp_1-1)+1,rank_0(B_1,ep_1)]$, and those where $B_1[i]=1$ are mapped to $B_2[sp_2',ep_2'] = B_2[z_1+rank_1(B_1,sp_1-1)+1,z_1+rank_1(B_1,ep_1)]$. 
The recursive process stops when the range becomes empty; when the sequence of highest bits makes the possible set of values either disjoint with  $[y_1,y_2]$ or included in $[y_1,y_2]$; or when we reach the last level. 
It can be shown that the recursion ends in $O(\log n)$ ranges, at most two per level, so that every value in those ranges is an answer. The corresponding $y$ values can be obtained by tracking them downwards as explained.

These data structures, and our results, hold in the RAM model with computer word size $w = \Theta(\log n)$. The wavelet matrix is then said to use $O(n)$ space---i.e., linear space---, which is counted in $w$-bit words. The wavelet matrix is easily built in $O(n\log n)$ time, and less~\cite{MNV16}.

Another relevant functionality that can be offered within $2n+o(n)$ bits is the so-called {\em range maximum query (RMQ)}: given a static array $A[1\dd n]$, we preprocess it in $O(n)$ time so that we can answer RMQs in $O(1)$ time~\cite{FH11}: $rmq(A,i,j)$ is a position $p$, $i\leq p \leq j$, such that $A[p] = \max \{ A[k],~ i \le k \le j \}$.  
The data structure does not need to maintain $A$. In this paper we will use RMQs where $A$ can undergo updates, see Sec.~\ref{sec:geom-data-structure}.

 
\subparagraph*{Suffix Arrays and Trees.}
The suffix tree~\cite{Weiner73} is a classic data structure on texts which is able to answer efficiently many different kinds of string processing queries~\cite{Gus97,Apo85}, which uses linear space and can be built in linear time~\cite{Weiner73,McCreight76,Farach97,Ukk95}. We give a brief recap; see Gusfield~\cite{Gus97} for more details. 

The suffix tree $\ST(T)$ of a text $T$ is the compact trie of the suffixes of $T$; it is a rooted tree whose edges are labeled by substrings of $T$ (stored as two pointers into $T$), and whose inner nodes are branching. The {\em label} $L(v)$ of a node $v$ is the concatenation of the labels of the edges on the root-to-$v$ path. There is a one-to-one correspondence between leaves and suffixes of $T$; $\leaf_i$ is then the unique leaf whose label equals the $i$th suffix $T[i\dd]$. The {\em stringdepth} $\sd(v)$ of a node $v$ is the length of its label, and we assume $\sd(v)$ is stored in $v$.

The suffix array $\SA$ of $T$ is a permutation of the index set $\{1,\ldots,n\}$ such that $\SA[i]=j$ if the $j$th suffix of $T$ is the $i$th in lexicographic order among all suffixes. The suffix array can be computed from the suffix tree, or directly from the text, in linear time and space~\cite{PST07,NZC11}. The inverse suffix array,%
\shorten{also a permutation of $\{1,\ldots,n\}$, is}
denoted $\ISA$,\shorten{and} can be computed in linear time using $\ISA[\SA[i]] = i$.


\section{The Lempel-Ziv (LZ) Parsing and its Bounded Version (\BLZ)}
\label{sec:lz}

The Lempel-Ziv (LZ) parsing of a text $T[1\dd n]$~\cite{LZ76}
produces a sequence of $z$ ``phrases'', which are substrings of $T$ whose concatenation is $T$. Each 
phrase is formed by the longest substring that has an occurrence 
starting earlier in $T$, plus the character that follows it. 

\begin{definition}
A {\em leftward parse} of $T[1\dd n]$ is a sequence of substrings $T[i\dd
i+\ell]$ (called {\em phrases}) whose concatenation is $T$ and such that there 
is an occurrence of each $T[i\dd i+\ell-1]$ starting before $i$ in $T$ (the
occurrence is called
the {\em source} of the phrase). The LZ parse of $T$ is the leftward parse of 
$T$ that, in a left-to-right process, chooses the longest possible phrases.
\end{definition}

The algorithm moves a pointer $i$
along $T$, from $i=1$ to $i=n$. 
At each step, the algorithm has already 
processed $T[1\dd i-1]$, and it must form the next phrase. As said, the phrase
is formed by (1) the longest prefix $T[i\dd i+\ell-1]$ of $T[i\dd ]$ that has
an occurrence in $T$ starting before position $i$, and (2) the next symbol 
$T[i+\ell]$. If $\ell>0$, then the occurrence of (1), $T[s\dd s+\ell-1] = 
T[i\dd i+\ell-1]$ with $s<i$, is called the source of $T[i\dd i+\ell-1]$.
Once suitable $s$ and $\ell$ have been determined, the next phrase is
$T[i\dd i+\ell]$ and the algorithm proceeds from $i \gets i+\ell+1$ onwards. 
The phrase $T[i\dd i+\ell]$ is encoded as the triple $(s,\ell,T[i+\ell])$, and 
if $\ell=0$ we can encode just the character $(T[i+\ell])$. 

This greedy parsing, which maximizes the phrase length at each step, turns
out to be optimal~\cite{LZ76}, that is, it produces the least
number $z$ of phrases among all the leftward parses of $T$.
Further, it can be computed in $O(n)$ time
\cite{RPE81,CPS08,OG11,KKP13,KKP13cpm,GB13,FIK15,KS16,BPK16,KKP16,FIKS18}. 

Note that phrases can overlap their sources, as sources must start---but not
necessarily end---before $i$. For example, the LZ parse of
$T=\mathtt{a}^{n-1}\mathtt{\$}$ is $(\mathtt{a})~(0,n-1,\mathtt{\$})$. For
illustrative purposes, we describe the parsings by writing bars,
``$\mathtt{|}$'',
between the formed phrases. The parsing of the example is then written as 
$\mathtt{a} | \mathtt{a}^{n-1}\mathtt{\$}$.
To illustrate the access problem, consider the LZ parsing of the text
$\mathtt{alabaralalabarda\$}$ (disregard for now the numbers below): 

\begin{center}
\begin{tabular}{c|c|cc|cc|cccc|ccccc|cc|}
    \a & \l & \a & \b & \a & \r & \a & \l & \a & \l & \a & \b & \a & \r & \d & \a & \texttt{\$}  \\
    0 & 0 & 1 & 0 & 1 & 0 & 1 & 1 & 2 & 0 & 2 & 1 & 2 & 1 & 0 & 1 & 0  \\
\end{tabular}
\end{center}

Assume we want to extract $T[11]=\mathtt{a}$. The position is the first of the
6th phrase, $\mathtt{abard}$, and it is copied from the third phrase,
$\mathtt{ab}$. In turn, the first position of that phrase is copied from the
first phrase, where $\mathtt{a}$ is stored in explicit form. We need then to
follow a {\em chain} of length two in order to extract $T[11]$, so
the length of that chain is the access cost. The numbers we wrote below
the symbols in the parse are the lengths of their chains.

\subparagraph*{Bounded Access Time Lempel-Ziv (\BLZ).}
We define a leftward parse we call Bounded Access Time Lempel-Ziv (\BLZ), which takes as a
parameter the maximum length $c$ any chain can have. A \BLZ\ parse is a leftward
parse where no chain is longer than $c$. Note that we do not require a \BLZ\ 
parse to be of minimal size. 
For example, a \BLZ\ parse for the above text with $c=1$ is as follows:

\begin{center}
\begin{tabular}{c|c|cc|cc|ccc|cc|cc|cc|cc|}
    \a & \l & \a & \b & \a & \r & \a & \l & \a & \l & \a & \b & \a & \r & \d & \a & \texttt{\$}  \\
    0 & 0 & 1 & 0 & 1 & 0 & 1 & 1 & 0 & 1 & 0 & 1 & 0 & 1 & 0 & 1 & 0  \\
\end{tabular}
\end{center}

When the LZ parse produces the phrase $T[i\dd i+\ell]$ from the source $T[s\dd
s+\ell-1]$ and the extra symbol $T[i+\ell]$, the character $T[i+\ell]$ is 
stored in explicit form, and thus its chain is of length zero. The chain
length of every other phrase symbol, $T[i+l]$ for $0 \le l < \ell$, is one
more than the chain length of its source symbol, $T[s+l]$.

A special case occurs when sources and targets overlap. If we want to extract 
$T[n-1]$ from $T=\mathtt{a}^{n-1}\mathtt{\$}$, we could note that it is copied
from $T[n-2]$, which is in turn copied from $T[n-3]$, and so on, implying a
chain of length $n-1$. Instead, we can note that our phrase $T[2\dd n]$ overlaps its source $T[1\dd n-2]$. In general, when the phrase $T[i\dd i+\ell-1]$ overlaps its source $T[s\dd s+\ell-1]$ by $0<b=i-s$ characters, this implies that the word $S = T[s\dd s+\ell-1] = T[i\dd i+\ell-1]$ has a {\em border} (a prefix which is also a suffix) of length $b$. It is well known that if $S$ has a border of length $b$, then $S$ has a period $p=|S|-b$, see~\cite[Ch.\ 8]{Lothaire2}. Therefore, $S$ can be written in the form $S = U^{\lfloor|S|/p\rfloor}V$, where $U$ is the $p$-length prefix of $S$ and $V$ a proper prefix of $U$, and thus, for all $l>p$, $S[l] = S[l \bmod p]$.

\begin{definition}[Chain length]\label{def:chainlength}
Let $T[i\dd i+\ell]$ be a phrase in a leftward parse of $T[1\dd n]$, whose
source is $T[s\dd s+\ell-1]$. The chain length of the explicit character is $C[i+\ell]=0$. If $\ell \leq i-s$ (i.e., there is no overlap between the source and the phrase), then for all $0\leq l<\ell$, $C[i+l] = C[s+l]+1$. Otherwise, for $0\leq l < i-s$, the chain length is $C[i+l] = C[s+l]+1$, and for $i-s \leq l < \ell$, the chain length is $C[i+l] = C[i + (l \bmod (i-s))]$. 
\end{definition}

We remark that a parsing like \BLZ\ is described as a baseline in the experimental 
results of one of the current authors' previous work~\cite{KNdcc10.1}, under the name LZ-Cost, 
but as no efficient parsing algorithm was devised for it, it could be tested only on the tiny texts of the
Canterbury Corpus ({\tt https://corpus.canterbury.ac.nz}). It also did not handle overlaps between sources and targets, so it did not perform well 
on the text $T=\mathtt{a}^n$. For testing the \BLZ\ parsing on large repetitive 
text collections we need an efficient parsing algorithm.


\section{A Greedy Parsing Algorithm for \BLZ}\label{sec:greedy}

In this section we describe an algorithm that, using $O(n)$ space and 
$O(n\log^3 n)$ time, produces a \BLZ\ parse of a text $T[1\dd n]$ by maximizing
the next phrase length at each step. We then show how to reduce the time to 
$O(n\log^2 n)$ at the price of increasing the space to $O(n\log n)$. Of course, unlike in LZ, this greedy algorithm does not in general produce an optimal \BLZ\ parse, since the problem is NP-hard.

\begin{definition}
A \BLZ\ parse of $T[1\dd n]$ with maximum chain length $c$ is a leftward parse 
of $T$ where the chain length of no position exceeds $c$. A {\em greedy \BLZ\ parse} 
is a \BLZ\ parse where each phrase, processed left to right, is as long as
possible.
\end{definition}

Let $T[1\dd i-1]$ be already processed. We call a prefix $T[i\dd i+\ell-1]$ of $T[i\dd]$ {\em valid} if $C[j]\leq c$ for all $j=i,\ldots,i+\ell-1$. A leftward parse of $T$ is therefore a \BLZ\ parse if and only if all phrases are valid. 
Our Greedy \BLZ\ parser proceeds then analogously to the original LZ parser. At 
each step, it has already processed $T[1\dd i-1]$, and it must find the next 
phrase, which is formed by (1) the longest valid 
prefix $T[i\dd i+\ell-1]$ of 
$T[i\dd]$ that has an occurrence $T[s\dd s+\ell-1]$ with $s<i$, and (2)
the next symbol $T[i+\ell]$. In other words, the algorithm enforces that 
every symbol in $T[s\dd s+\ell-1]$ must have a chain length less than $c$,
the maximum chain length allowed. The phrase $T[i\dd i+\ell]$ is encoded just as in the standard LZ, as a triple
$(s,\ell,T[i+\ell])$. 

To efficiently find $s$ and $\ell$, our \BLZ\ parsing algorithm 
stores the following structures:
\begin{enumerate}
\item The suffix array $\SA[1\dd n]$ of $T$, represented as a wavelet matrix
\cite{CNO15}.
\item The inverse suffix array $\ISA[1\dd n]$ of $T$, represented in plain form.
\item An array $C[1\dd n]$, where $C[i]$ is the chain length of $i$.
Note that $C[i]$ is defined only for the already parsed positions of $T$.
\item An array $D[1\dd n]$, where $D[s]$ is the minimum $d \ge 0$ such that
$C[s+d] = c$. If no such a $d$ exists (in particular, because $C[i]$ is
defined only for the parsed prefix), then $D[s] = \infty$ (which holds 
initially for all $s$). 
\item For each level of the wavelet matrix of $\SA$, a special {\em dynamic 
RMQ} data structure to track the text positions that can be used. This
structure is related to the values of $D$ and therefore it changes along the
parsing.
\end{enumerate}

Note that the definition of \BLZ\ implies that, if the source of 
$T[i\dd i+\ell-1]$ is $T[s\dd s+\ell-1]$, then it must be that $\ell \le D[s]$. This motivates the following observation: 

\begin{observation}
Let $T[1\dd i-1]$ be already processed. A prefix $T[i\dd i+\ell-1]$ of $T[i\dd ]$ is valid if and only if there exists a source $T[s\dd s+\ell-1]$ such that 
\begin{enumerate}[(i)]
\item its lexicographic position satisfies $\ISA[s] \in [sp\dd ep]$, where
$[sp\dd ep]$ is the suffix array range of $T[i\dd i+\ell-1]$ (i.e.,
$T[s\dd s+\ell-1] = T[i\dd i+\ell-1]$); 
\item its starting position in $T$ is $s < i$; and 
\item it does not use forbidden text positions, that is, $\ell \le D[s]$.
\end{enumerate}
\end{observation}

The parsing then must find the longest valid prefix $T[i\dd i+\ell-1]$ of
$T[i\dd ]$. We do so by testing the consecutive values $\ell=1,2,\ldots$.
Note that, once we have determined the next phrase $T[i\dd i+\ell]$, we must
update $C$ and $D$ as follows: (1) 
 $C[i+l] \gets C[s+l]+1$ for all $0 \le l < \ell$, and $C[i+\ell] \gets
0$.\footnote{Recall that a special case occurs if $T[i\dd i+\ell-1]$ overlaps
$T[s\dd s+\ell-1]$: we start copying from $k=s$ and increasing $k$ and, whenever 
$k = s+l = i$, we restart copying from $k=s$.}, and (2) 
Every time we obtain $C[t]=c$ in the previous point, we set $D[k] \gets t-k$
for all $k' < k \le t$, where $k'$ is the last position where $D[k'] <
\infty$ (so $k'=0$ in the beginning and we reset $k' \gets t$ after this
process). 

\begin{figure}[t]
    \centering
    \resizebox{0.5\textwidth}{!}{
    \tikzset{every picture/.style={line width=0.75pt}} 
    
    \begin{tikzpicture}[x=0.75pt,y=0.75pt,yscale=-1,xscale=1]
    
    \draw  [dash pattern={on 4.5pt off 4.5pt}] (53.33,38) -- (172,38) -- (172,151.67) -- (53.33,151.67) -- cycle ;
    \draw  [dash pattern={on 4.5pt off 4.5pt}]  (172,152) -- (310,256) ;
    \draw  [dash pattern={on 4.5pt off 4.5pt}]  (172,38) -- (310,142) ;
    \draw  [dash pattern={on 4.5pt off 4.5pt}]  (54,152) -- (192,256) ;
    \draw    (118,33) -- (118,43) ;
    \draw    (49,72) -- (57,72) ;
    \draw    (49,104) -- (57,104) ;
    \draw  [dash pattern={on 0.84pt off 2.51pt}]  (54,72) -- (118,72) ;
    \draw  [dash pattern={on 0.84pt off 2.51pt}]  (54,104) -- (118,104) ;
    \draw    (198,52) -- (198,62) ;
    \draw  [dash pattern={on 0.84pt off 2.51pt}]  (54,104) -- (80,124) ;
    \draw  [dash pattern={on 0.84pt off 2.51pt}]  (54,72) -- (80,92) ;
    \draw    (144,124) -- (284,230) ;
    \draw  [dash pattern={on 0.84pt off 2.51pt}]  (118,38) -- (118,104) ;
    \draw  [dash pattern={on 0.84pt off 2.51pt}]  (143,91) -- (118,72) ;
    \draw    (144,92) -- (144,124) ;
    \draw  [dash pattern={on 0.84pt off 2.51pt}]  (144,124) -- (119,105) ;
    \draw    (80,92) -- (144,92) ;
    \draw    (80,92) -- (80,124) ;
    \draw    (80,124) -- (144,124) ;
    \draw    (80,124) -- (220,230) ;
    \draw    (80,92) -- (220,198) ;
    \draw    (144,92) -- (284,198) ;
    \draw    (220,198) -- (220,230) ;
    \draw    (220,198) -- (284,198) ;
    \draw    (284,198) -- (284,230) ;
    \draw    (220,230) -- (284,230) ;
    \draw  [dash pattern={on 4.5pt off 4.5pt}]  (192,256) -- (310,256) ;
    \draw  [dash pattern={on 4.5pt off 4.5pt}]  (310,142) -- (310,256) ;
    \draw  [dash pattern={on 0.84pt off 2.51pt}]  (144,58) -- (198,58) ;
    \draw  [dash pattern={on 0.84pt off 2.51pt}]  (144,58) -- (144,92) ;
    
    \draw (73,4) node [anchor=north west][inner sep=0.75pt]  [font=\small] [align=left] {Text positions};
    \draw (26,44) node [anchor=north west][inner sep=0.75pt]  [font=\small,rotate=-90] [align=left] {$\SA$ positions};
    \draw (115,20) node [anchor=north west][inner sep=0.75pt]  [font=\footnotesize] [align=left] {$i$};
    \draw (47,63) node [anchor=north west][inner sep=0.75pt]  [rotate=-90] [align=left] {$sp$};
    \draw (47,96) node [anchor=north west][inner sep=0.75pt]  [rotate=-90] [align=left] {$ep$};
    \draw (194,38) node [anchor=north west][inner sep=0.75pt]  [font=\footnotesize] [align=left] {$\ell $};
    \draw (132.41,66.18) node [anchor=north west][inner sep=0.75pt]  [font=\footnotesize,rotate=-38.96] [align=left] {$\ell $};
    \draw (194.96,12.85) node [anchor=north west][inner sep=0.75pt]  [font=\small,rotate=-36.7] [align=left] {$D$ values (undergo updates)};

    \end{tikzpicture}
    }
    
    \caption{General scheme of our translation of queries onto a 3-dimensional data structure.}
    \label{fig:3d-DS}
\end{figure}
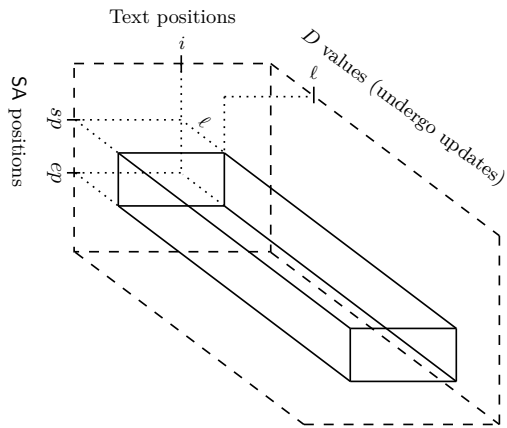

Note that points (i) and (ii) above correspond to the classic LZ parsing 
problem. In particular, they correspond to determining whether there are points
in the range $[sp,ep] \times [1,i-1]$ of the grid represented by our 
wavelet matrix, which represents the points $(j,\SA[j])$. As the wavelet
matrix answers this query in time $O(\log n)$, this yields an $O(n\log n)$ LZ 
parsing algorithm.\shorten{\footnote{\shorten{As mentioned, linear-time algorithms exist, but we
do not know how to extend them to \BLZ.}}} Point (iii),
however, is exclusive to \BLZ. It can be handled by converting the grid into 
a three-dimensional mesh, where we store the values $(j,\SA[j],D[\SA[j]])$ and
look for the existence of points in the range $[sp,ep] \times [1,i-1] \times
[\ell,n]$. Note that we need to determine whether the range is empty and, if
it is not, retrieve a point from it (whose second coordinate is the desired 
$s$). In addition, as the array $D$ is modified along the parsing, we 
need a dynamic 3-dimensional data structure: every time we modify $D$ in point
2 above, our data structure changes (this occurs up to $n$ times).
See Fig.~\ref{fig:3d-DS}.

Our 3-dimensional problem, then, (a) is essentially a range emptiness query
(where we must return one point if there are any), (b) the search is 4-sided 
(though our solution handles 5-sided queries), and (c) the updates in $D$ occur
only to convert some $D[k]= \infty$ into a smaller value, so each value $D[k]$ 
changes at most once along the parsing process (yet, our solution handles 
arbitrary updates along the coordinate where the query is one-sided). 
We have found no linear-space solutions to this problem in the literature; 
only solutions to less general ones or using super-linear space (indeed, more
than $O(n\log n)$):
(1) linear space for {\em two dimensions}, with $O(\log n)$ query time and 
$O(\log^{3+\epsilon} n)$ update time~\cite{Nek09}; (2) linear space for three
dimensions with {\em no updates}, with $O(\log n / \log\log n)$ query time
\cite{CNRT22}; (3) {\em super-linear} space (at least $O(n\log^{1.33} n)$ for 
three dimensions), with $O((\log n/\log\log n)^2)$ query time and
$O(\log^{1.33+\epsilon} n)$ update time~\cite{CT18}.
In the next section we describe our data structures for this problem: 
one uses linear space and $O(\log^3 n)$ query and update time; the other uses 
$O(n\log n)$ space and $O(\log^2 n)$ query time. This yields our first main result. 

\begin{theorem}
A Greedy \BLZ\ parse of a text $T[1\dd n]$ can be computed using $O(n)$ 
space and $O(n\log^3 n)$ time, or $O(n\log n)$ space and $O(n\log^2 n)$ time.
\end{theorem}


\section{A Geometric Data Structure}\label{sec:geom-data-structure}

To solve the 3-dimensional search problem we associate, with each
level of the wavelet matrix, a data structure that represents the sequence of 
values $D[k]$ in the order the text positions $k$ are listed in that level.
Because in linear space we cannot store the actual values in every wavelet
matrix level, we store only a dynamic RMQ data structure on the internal
levels, and store the explicit values only in (the order
corresponding to) the last level (in a wavelet matrix, that final
level is not the text order, thus we need another array to map it to $D$).

\begin{figure}
    \centering
    \resizebox{\textwidth}{!}{
    \tikzset{every picture/.style={line width=0.75pt}} 
    
    \begin{tikzpicture}[x=0.75pt,y=0.75pt,yscale=-1,xscale=1]
    
    \draw   (26,22) -- (292,22) -- (292,32) -- (26,32) -- cycle ;
    \draw   (26,92) -- (158,92) -- (158,102) -- (26,102) -- cycle ;
    \draw   (26,162) -- (86,162) -- (86,172) -- (26,172) -- cycle ;
    \draw   (26,34) -- (292,34) -- (292,38) -- (26,38) -- cycle ;
    \draw    (88,18) -- (88,26) ;
    \draw    (198,18) -- (198,26) ;
    \draw   (26,104) -- (292,104) -- (292,108) -- (26,108) -- cycle ;
    \draw   (26,174) -- (292,174) -- (292,178) -- (26,178) -- cycle ;
    \draw   (160,92) -- (292,92) -- (292,102) -- (160,102) -- cycle ;
    \draw    (56,88) -- (56,96) ;
    \draw    (120,88) -- (120,96) ;
    \draw    (124,42) -- (86,90) ;
    \draw    (194,42) -- (226,90) ;
    \draw   (88,162) -- (156,162) -- (156,172) -- (88,172) -- cycle ;
    \draw    (106,159) -- (106,167) ;
    \draw    (138,159) -- (138,167) ;
    \draw    (100,111) -- (128,154) ;
    \draw    (82,111) -- (52,154) ;
    \draw   (158,162) -- (228,162) -- (228,172) -- (158,172) -- cycle ;
    \draw   (230,162) -- (292,162) -- (292,172) -- (230,172) -- cycle ;
    \draw    (240,112) -- (268,155) ;
    \draw    (222,112) -- (192,155) ;
    \draw    (106,174) -- (106,178) ;
    \draw    (138,174) -- (138,178) ;
    \draw [color={rgb, 255:red, 208; green, 2; blue, 27 }  ,draw opacity=1 ]  (120,176) .. controls (120,174.9) and (120.9,174) .. (122,174) .. controls (123.1,174) and (124,174.9) .. (124,176) .. controls (124,177.1) and (123.1,178) .. (122,178) .. controls (120.9,178) and (120,177.1) .. (120,176) -- cycle ;
    \draw    (158,188) .. controls (134.64,188.86) and (127.42,183.8) .. (123.63,181.1) ;
    \draw [shift={(122,180)}, rotate = 29.74] [color={rgb, 255:red, 0; green, 0; blue, 0 }  ][line width=0.75]    (10.93,-3.29) .. controls (6.95,-1.4) and (3.31,-0.3) .. (0,0) .. controls (3.31,0.3) and (6.95,1.4) .. (10.93,3.29)   ;
    \draw   (26,259) -- (292,259) -- (292,269) -- (26,269) -- cycle ;
    \draw    (96,259) -- (96,269) ;
    \draw    (104,259) -- (104,269) ;
    \draw    (136,280) .. controls (108.54,279.24) and (109.7,281.37) .. (101.4,273.35) ;
    \draw [shift={(100,272)}, rotate = 44.12] [color={rgb, 255:red, 0; green, 0; blue, 0 }  ][line width=0.75]    (10.93,-3.29) .. controls (6.95,-1.4) and (3.31,-0.3) .. (0,0) .. controls (3.31,0.3) and (6.95,1.4) .. (10.93,3.29)   ;
    \draw   (88,200) -- (120,200) -- (120,210) -- (88,210) -- cycle ;
    \draw   (122,200) -- (156,200) -- (156,210) -- (122,210) -- cycle ;
    \draw   (88,221) -- (102,221) -- (102,231) -- (88,231) -- cycle ;
    \draw   (104,221) -- (120,221) -- (120,231) -- (104,231) -- cycle ;
    \draw   (88,244) -- (94,244) -- (94,254) -- (88,254) -- cycle ;
    \draw   (96,244) -- (104,244) -- (104,254) -- (96,254) -- cycle ;
    \draw  [color={rgb, 255:red, 208; green, 2; blue, 27 }  ,draw opacity=1 ] (105,205) .. controls (105,203.9) and (105.9,203) .. (107,203) .. controls (108.1,203) and (109,203.9) .. (109,205) .. controls (109,206.1) and (108.1,207) .. (107,207) .. controls (105.9,207) and (105,206.1) .. (105,205) -- cycle ;
    \draw [color={rgb, 255:red, 208; green, 2; blue, 27 }  ,draw opacity=1 ]  (95,226) .. controls (95,224.9) and (95.9,224) .. (97,224) .. controls (98.1,224) and (99,224.9) .. (99,226) .. controls (99,227.1) and (98.1,228) .. (97,228) .. controls (95.9,228) and (95,227.1) .. (95,226) -- cycle ;
    \draw [color={rgb, 255:red, 208; green, 2; blue, 27 }  ,draw opacity=1 ]  (98,249) .. controls (98,247.9) and (98.9,247) .. (100,247) .. controls (101.1,247) and (102,247.9) .. (102,249) .. controls (102,250.1) and (101.1,251) .. (100,251) .. controls (98.9,251) and (98,250.1) .. (98,249) -- cycle ;
    \draw    (124,186) -- (134,198) ;
    \draw    (118,186) -- (108,198) ;
    \draw    (102,212) -- (96,218) ;
    \draw    (108,212) -- (114,218) ;
    \draw    (94,234) -- (92,242) ;
    \draw    (99,234) -- (101,242) ;
    \draw   (80.58,196.33) .. controls (75.91,196.32) and (73.57,198.64) .. (73.56,203.31) -- (73.53,215.98) .. controls (73.51,222.65) and (71.17,225.97) .. (66.5,225.96) .. controls (71.17,225.97) and (73.49,229.31) .. (73.47,235.98)(73.48,232.98) -- (73.44,248.65) .. controls (73.43,253.32) and (75.75,255.66) .. (80.42,255.67) ;
    \draw    (526,37.07) -- (602,91.07) ;
    \draw   (380,173.07) -- (676,173.07) -- (676,184.07) -- (380,184.07) -- cycle ;
    \draw    (526,37.07) -- (454,91.07) ;
    \draw    (454,91.07) -- (492,122.07) ;
    \draw    (454,91.07) -- (418,121.07) ;
    \draw    (491,121.07) -- (510,147.07) ;
    \draw    (491,121.07) -- (472,147.07) ;
    \draw    (472,147.07) -- (484,172.07) ;
    \draw    (472,147.07) -- (460,172.07) ;
    \draw    (510,147.07) -- (522,172.07) ;
    \draw    (510,147.07) -- (498,172.07) ;
    \draw    (418,121.07) -- (437,147.07) ;
    \draw    (418,121.07) -- (399,147.07) ;
    \draw    (399,147.07) -- (414,173.07) ;
    \draw    (399,147.07) -- (384,173.07) ;
    \draw    (437,147.07) -- (450,172.07) ;
    \draw    (437,147.07) -- (422,173.07) ;
    \draw    (602,91.07) -- (638,121.07) ;
    \draw    (602,91.07) -- (566,121.07) ;
    \draw    (638,121.07) -- (657,147.07) ;
    \draw    (638,121.07) -- (619,147.07) ;
    \draw    (619,147.07) -- (630,172.07) ;
    \draw    (619,147.07) -- (608,172.07) ;
    \draw    (657,147.07) -- (670,172.07) ;
    \draw    (657,147.07) -- (646,172.07) ;
    \draw    (566,121.07) -- (585,147.07) ;
    \draw    (566,121.07) -- (547,147.07) ;
    \draw    (547,147.07) -- (560,172.07) ;
    \draw    (547,147.07) -- (534,172.07) ;
    \draw    (585,147.07) -- (596,172.07) ;
    \draw    (585,147.07) -- (574,172.07) ;
    \draw  [color={rgb, 255:red, 208; green, 2; blue, 27 }  ,draw opacity=1 ] (444,179.07) .. controls (444,176.31) and (446.24,174.07) .. (449,174.07) .. controls (451.76,174.07) and (454,176.31) .. (454,179.07) .. controls (454,181.83) and (451.76,184.07) .. (449,184.07) .. controls (446.24,184.07) and (444,181.83) .. (444,179.07) -- cycle ;
    \draw  [color={rgb, 255:red, 208; green, 2; blue, 27 }  ,draw opacity=1 ] (480,179.07) .. controls (480,176.31) and (482.24,174.07) .. (485,174.07) .. controls (487.76,174.07) and (490,176.31) .. (490,179.07) .. controls (490,181.83) and (487.76,184.07) .. (485,184.07) .. controls (482.24,184.07) and (480,181.83) .. (480,179.07) -- cycle ;
    \draw  [color={rgb, 255:red, 208; green, 2; blue, 27 }  ,draw opacity=1 ] (556,179.07) .. controls (556,176.31) and (558.24,174.07) .. (561,174.07) .. controls (563.76,174.07) and (566,176.31) .. (566,179.07) .. controls (566,181.83) and (563.76,184.07) .. (561,184.07) .. controls (558.24,184.07) and (556,181.83) .. (556,179.07) -- cycle ;
    \draw  [color={rgb, 255:red, 208; green, 2; blue, 27 }  ,draw opacity=1 ] (602,179.07) .. controls (602,176.31) and (604.24,174.07) .. (607,174.07) .. controls (609.76,174.07) and (612,176.31) .. (612,179.07) .. controls (612,181.83) and (609.76,184.07) .. (607,184.07) .. controls (604.24,184.07) and (602,181.83) .. (602,179.07) -- cycle ;
    \draw  [color={rgb, 255:red, 208; green, 2; blue, 27 }  ,draw opacity=1 ] (387,253.07) .. controls (387,250.31) and (389.24,248.07) .. (392,248.07) .. controls (394.76,248.07) and (397,250.31) .. (397,253.07) .. controls (397,255.83) and (394.76,258.07) .. (392,258.07) .. controls (389.24,258.07) and (387,255.83) .. (387,253.07) -- cycle ;
    \draw  [color={rgb, 255:red, 208; green, 2; blue, 27 }  ,draw opacity=1 ] (487.06,166.43) -- (482.22,164.25) -- (483.7,163.49) -- (477.48,151.31) -- (480.43,149.8) -- (486.66,161.98) -- (488.13,161.23) -- cycle ;
    \draw   (474.17,101.55) -- (468.86,101.36) -- (469.95,100.1) -- (459.59,91.17) -- (461.76,88.66) -- (472.12,97.59) -- (473.2,96.33) -- cycle ;
    \draw  [color={rgb, 255:red, 208; green, 2; blue, 27 }  ,draw opacity=1 ] (473.51,135.87) -- (473.76,130.57) -- (475.01,131.67) -- (484.06,121.41) -- (486.55,123.61) -- (477.5,133.86) -- (478.74,134.96) -- cycle ;
    \draw   (506.84,46.03) -- (508.16,40.89) -- (509.16,42.21) -- (520.1,34) -- (522.09,36.65) -- (511.15,44.87) -- (512.15,46.2) -- cycle ;
    \draw   (385.24,161.22) -- (384.8,155.93) -- (386.18,156.85) -- (393.81,145.5) -- (396.56,147.35) -- (388.93,158.7) -- (390.31,159.63) -- cycle ;
    \draw   (422.76,161.92) -- (422.31,156.63) -- (423.69,157.55) -- (431.32,146.2) -- (434.07,148.05) -- (426.44,159.41) -- (427.82,160.33) -- cycle ;
    \draw   (495.14,166.37) -- (494.4,161.12) -- (495.83,161.96) -- (502.8,150.2) -- (505.66,151.89) -- (498.68,163.66) -- (500.11,164.5) -- cycle ;
    \draw  [color={rgb, 255:red, 208; green, 2; blue, 27 }  ,draw opacity=1 ] (550,134.21) -- (549.99,128.91) -- (551.28,129.94) -- (559.81,119.25) -- (562.41,121.32) -- (553.88,132.01) -- (555.18,133.05) -- cycle ;
    \draw   (622.29,133.93) -- (622.28,128.62) -- (623.58,129.65) -- (632.11,118.96) -- (634.7,121.03) -- (626.17,131.72) -- (627.47,132.76) -- cycle ;
    \draw   (434.52,133.96) -- (429.5,133.17) -- (430.77,132.09) -- (421.74,121.44) -- (424.27,119.29) -- (433.3,129.95) -- (434.56,128.87) -- cycle ;
    \draw  [color={rgb, 255:red, 208; green, 2; blue, 27 }  ,draw opacity=1 ] (562.34,166.05) -- (557.36,164.21) -- (558.78,163.35) -- (551.73,151.63) -- (554.58,149.92) -- (561.62,161.64) -- (563.05,160.79) -- cycle ;
    \draw   (599.06,166.43) -- (594.22,164.25) -- (595.7,163.49) -- (589.48,151.31) -- (592.43,149.8) -- (598.66,161.98) -- (600.13,161.23) -- cycle ;
    \draw   (632.96,164.56) -- (628.12,162.38) -- (629.59,161.62) -- (623.37,149.44) -- (626.33,147.93) -- (632.55,160.11) -- (634.03,159.36) -- cycle ;
    \draw   (644.2,163.24) -- (643.32,158) -- (644.77,158.81) -- (651.45,146.88) -- (654.35,148.5) -- (647.67,160.43) -- (649.11,161.24) -- cycle ;
    \draw   (620.82,100.59) -- (615.52,100.39) -- (616.61,99.14) -- (606.26,90.19) -- (608.43,87.68) -- (618.78,96.62) -- (619.86,95.37) -- cycle ;
    \draw  [color={rgb, 255:red, 74; green, 144; blue, 226 }  ,draw opacity=1 ] (449,174.07) -- (454,179.07) -- (449,184.07) -- (444,179.07) -- cycle ;
    \draw  [color={rgb, 255:red, 74; green, 144; blue, 226 }  ,draw opacity=1 ] (491,117.07) -- (496,122.07) -- (491,127.07) -- (486,122.07) -- cycle ;
    \draw  [color={rgb, 255:red, 74; green, 144; blue, 226 }  ,draw opacity=1 ] (567,117.07) -- (572,122.07) -- (567,127.07) -- (562,122.07) -- cycle ;
    \draw  [color={rgb, 255:red, 74; green, 144; blue, 226 }  ,draw opacity=1 ] (607,174.07) -- (612,179.07) -- (607,184.07) -- (602,179.07) -- cycle ;
    \draw  [color={rgb, 255:red, 74; green, 144; blue, 226 }  ,draw opacity=1 ] (392,226.07) -- (397,231.07) -- (392,236.07) -- (387,231.07) -- cycle ;
    \draw   (440,184.73) .. controls (440,189.4) and (442.33,191.73) .. (447,191.73) -- (517.42,191.73) .. controls (524.09,191.73) and (527.42,194.06) .. (527.42,198.73) .. controls (527.42,194.06) and (530.75,191.73) .. (537.42,191.73)(534.42,191.73) -- (607.83,191.73) .. controls (612.5,191.73) and (614.83,189.4) .. (614.83,184.73) ;
    
    \draw (81,1) node [anchor=north west][inner sep=0.75pt]  [font=\footnotesize] [align=left] {$sp_1$};
    \draw (191,2) node [anchor=north west][inner sep=0.75pt]  [font=\footnotesize] [align=left] {$ep_1$};
    \draw (294,30) node [anchor=north west][inner sep=0.75pt]  [font=\footnotesize] [align=left] {RMQ on $D_{1}$};
    \draw (294,100) node [anchor=north west][inner sep=0.75pt]  [font=\footnotesize] [align=left] {RMQ on $D_{2}$};
    \draw (294,170) node [anchor=north west][inner sep=0.75pt]  [font=\footnotesize] [align=left] {RMQ on $D_{3}$};
    \draw (49,71) node [anchor=north west][inner sep=0.75pt]  [font=\footnotesize] [align=left] {$sp_{2}$};
    \draw (113,71) node [anchor=north west][inner sep=0.75pt]  [font=\footnotesize] [align=left] {$ep_{2}$};
    \draw (99,142) node [anchor=north west][inner sep=0.75pt]  [font=\footnotesize] [align=left] {$sp_{3}$};
    \draw (131,142) node [anchor=north west][inner sep=0.75pt]  [font=\footnotesize] [align=left] {$ep_{3}$};
    \draw (158,181) node [anchor=north west][inner sep=0.75pt]  [font=\footnotesize] [align=left] {candidate pos};
    \draw (136,273) node [anchor=north west][inner sep=0.75pt]  [font=\footnotesize] [align=left] {candidate value};
    \draw (47,225) node  [font=\footnotesize] [align=left] {\begin{minipage}[lt]{39.44pt}\setlength\topsep{0pt}
    track position along \\wavelet matrix
    \end{minipage}};
    \draw (294,254) node [anchor=north west][inner sep=0.75pt]   [align=left] {$D$};
    \draw (513,198.07) node [anchor=north west][inner sep=0.75pt]  [font=\footnotesize] [align=left] {max};
    \draw (679,172.07) node [anchor=north west][inner sep=0.75pt]   [align=left] {$D_{3}$};
    \draw (399,246.07) node [anchor=north west][inner sep=0.75pt]  [font=\small] [align=left] {= candidates to max};
    \draw (653,64.07) node [anchor=north west][inner sep=0.75pt]   [align=left] {$H_{3}$};
    \draw (399,224.07) node [anchor=north west][inner sep=0.75pt]  [font=\small] [align=left] {= roots of subtrees covering the query interval};

    \end{tikzpicture}
    }
    
    \caption{On the left, we reach a candidate area $[sp_3,ep_3]$ of the wavelet matrix and must obtain its maximum $D$ value using the (dynamic) RMQ data structure for $D_3$. The tree $H_3$ for this RMQ structure is shown on the right. Arrows point to the child holding the maximum value in $D_3$. Blue diamonds are the roots $v_3^1,\ldots,v_3^4$ of the subtrees that cover the query area $[sp_3,ep_3]$ and red circles are the candidates in the range. The left plot shows how we find the actual value of one of those circles by tracking it down in the wavelet matrix.}
    \label{fig:wm-rmq}
\end{figure}
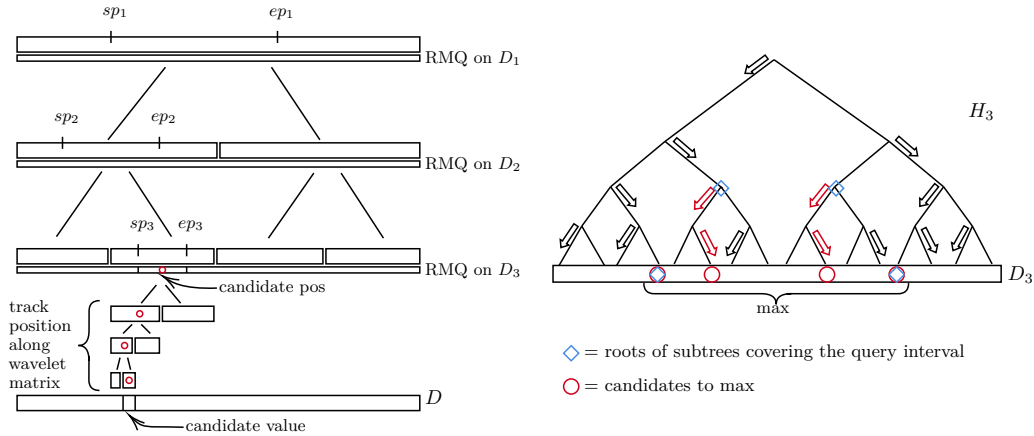

Let $D_l$ be the array $D$ permuted in the way it corresponds to level $l$ of 
the wavelet matrix. 
The dynamic RMQ structure for level $l$ is then a 
heap-shaped perfectly balanced tree $H_l[1\dd n]$ whose leaves (implicitly) point
to the entries of $D_l$. 
The nodes $H_l[p]$ store only one bit, $0$ indicating that the maximum in the 
subtree is to the left and $1$ indicating that it is to the right. By
navigating $H_l$ from the root $p$ of any subtree, moving to $H_l[2p]$
if $H_l[p]=0$ and $H_l[2p+1]$ if $H_l[p]=1$, we arrive in $O(\log n)$ time at
the position $p$ where $D_l[p]$ is maximum below that subtree. The actual
value $D_l[p]$ is obtained in other $O(\log n)$ time by tracking position $p$
downwards in the wavelet matrix, from level $l$ until the last
level, where the values of $D$ are explicitly stored. See Fig.~\ref{fig:wm-rmq} (right); ignore the query for now.

\subparagraph*{Updates.}
When a value $D[k]$ decreases from $\infty$, we obtain its position in the 
top-level of the wavelet matrix as $p = \ISA[k]$; thus we must reflect in
$H_1$ the decrease in the
value of $D_1[p]$. By halving $p$ successively we arrive at its ancestors,
$H_1[p_h]$ for $p_h = \lfloor p/2^h\rfloor$, $h=1, 2, \ldots$ We traverse the
path upwards, recomputing the maximum value $m$ below $p_h$ and modifying
accordingly the bits of $H_1[p_h]$. Initially, this new maximum is 
$m = D_1[p] = D[k]$. At any point in the traversal, if the parent 
$H_1[p_h]$ of the current node indicates that the maximum below $p_h$ descends 
from the {\em other} child of $p_h$, then we can stop updating of $H_1$, because
decreasing $D_1[p]$ does not require further changes. Otherwise, we must
obtain the maximum value $m'$ below the other child of $H_1[p_h]$ and compare it
with $m$. The value $m'$ is obtained in $O(\log n)$ time as explained in the 
previous paragraph. We set $H_1[p_h]$ depending on which is larger between $m$
and $m'$, update $m \gets \max(m,m')$, and continue upwards. This process
takes $O(\log^2 n)$ time as we traverse all the levels of $H_1$. We then track
position $p$ downwards to the second level of the wavelet matrix, update $H_2$
in the same way, and continue updating $H_l$ on all the wavelet matrix levels
$l$, for a total update time of $O(\log^3 n)$.

\subparagraph*{Searches.} The search for a range $[sp,ep] \times [1,i-1]
\times [\ell,n]$ first determines, as in the normal wavelet matrix search
algorithm, the $O(\log n)$ maximal ranges that cover $[1,i-1]$ along the 
wavelet matrix levels $l$ (there is at most one range per level because the
range $[1,i-1]$ is one-sided; otherwise there could be two), and maps $[sp,ep]$
to $[sp_l,ep_l]$ on each such range (see Sec.~\ref{sec:basics}), all in time $O(\log n)$. We then need to 
determine if there is some value $D_l[p] \ge \ell$ below some of the ranges
$[sp_l,ep_l]$ (see the top-left part of Fig.~\ref{fig:wm-rmq}). Each such range is then, again, decomposed into $O(\log n)$
maximal nodes $v_l^1, v_l^2, \ldots$ of $H_l$ (see the right of Fig.~\ref{fig:wm-rmq}). We find, in $O(\log n)$ time,
the maximum value of $D_l$ below each node $v_l^j$, stopping as soon as we
find some value $\ge \ell$. Note that we use $O(\log n)$ time to find the {\em position} of the maximum in $D_l$ using $H_l$, and then $O(\log n)$ time to find the {\em value} of that maximum by tracking the position down in the wavelet matrix (see the bottom left of Fig.~\ref{fig:wm-rmq}). Since we have $O(\log n)$ ranges $[sp_l,ep_l]$, each yielding $O(\log n)$ candidates $v_l^i$, and the maximum of each candidate is computed in $O(\log n)$ time, the whole search process takes time $O(\log^3 n)$.

\subparagraph*{Generalizations.} Though not necessary for our problem, we remark
that our update process can be extended to arbitrary updates on the 
third coordinate, $D[k]$, not only to reductions in value. Further, our search
could support five-sided ranges, not only four-sided, because we would still
have $O(\log n)$ ranges $[sp_l,ep_l]$ if the range of the second coordinate
was two-sided. Only the range of the third coordinate (the one supporting the 
updates) must be one-sided.

\subparagraph*{Faster and larger.} By storing the values of $D_l$ in each node 
of $H_l$ for each wavelet matrix level $l$, the space increases to $O(n \log n)$
but the time of updates and searches decreases to $O(\log^2 n)$, as we have
now the maximum below any $H_l[p_h]$ readily available in $O(1)$ time.


\section{The Minmax Parsing Algorithm}\label{sec:minmax}

We note that our Greedy \BLZ\ algorithm does not necessarily produce the smallest greedy parse, because it may fail in choosing the best {\em source} for the longest phrase.
Consider, say, the text $T=\mathtt{alabaralalabarda\$}$ and $c=2$.
Our implementation parses it into $8$ phrases as

\begin{center}
\begin{tabular}{c|c|cc|cc|cccc|ccc|cc|cc|}
    \a & \l & \a & \b & \a & \r & \a & \l & \a & \l & \a & \b & \a & \r & \d & \a & \texttt{\$}  \\
    0 & 0 & 1 & 0 & 2 & 0 & 1 & 1 & 2 & 0 & 2 & 1 & 0 & 1 & 0 & 1 & 0  \\
\end{tabular}
\end{center}

\noindent because it chooses $T[3]$ as the source for the 4th phrase, {\tt ar}, and then
$T[5]$ has a chain of length two and cannot be used again. If,
instead, we choose $T[1]$ as the source of the 4th phrase, the 
chain of $T[5]$ will be of length $1$ and we could parse $T$ into $7$
phrases, just as the first parse shown in Sec.~\ref{sec:lz}.

Our second algorithm, the Minmax parser,  always chooses a source that minimizes the maximum chain length in the phrase, among all possible sources. It compromises however on the {\em length} of the phrase, by not always choosing the longest admissible phrase. As we will see, this is well worth it: Minmax always produces a much better compression than Greedy.

\subparagraph*{High-level description of the Minmax parser.} Let $T[1\dd i-1]$ be already processed. We will call a prefix $T[i\dd i+\ell-1]$ of $T[i\dd]$ {\em admissible} if it has a source $T[s\dd s+\ell-1]$ with $\max C[s\dd s+\ell-1] < c$. 
We would ideally like to find the longest admissible prefix of $T[i\dd]$, and then choose its best source if there is more than one. We will use an enhanced suffix tree of the text; this will allow us to store additional information in the nodes. Navigating in the suffix tree, we will then be able to choose the longest admissible prefix {\em which ends in some node} (i.e., not necessarily the longest), and then choose the best source of this prefix. In order to do this, we will match the current suffix $T[i\dd]$ in the usual way in the suffix tree, using the desired information written in the nodes. As this information is dynamic, however, we will have to update it during the algorithm. The algorithm thus proceeds by (1) matching the suffix $T[i\dd]$ in the suffix tree and returning the next phrase and its source, and (2) updating the annotation.

\subparagraph*{Annotation of the suffix tree.} 
On the suffix tree of $T$, we annotate each node $v$ with three variables $\minmax(v), \txtpos(v)$, and a Boolean $\real(v)$, initializing $\minmax(v)$ to $+\infty$, $\txtpos(v)$ to $-1$, and $\real(v)$ to $0$. Recall that $L(v)$ is the label of $v$ and $\sd(v)$ its length. The variables $\minmax(v)$ and $\txtpos(v)$ will point to the current best candidate of an occurrence of $L(v)$, with $\txtpos(v)$ its starting position and $\minmax(v)$ the maximum $C$-value within this occurrence. The Boolean $\real(v)$ indicates whether this value is {\em realistic} ($\real(v)=1$), i.e., a full occurrence with this value has already been seen, or only {\em optimistic} ($\real(v)=0$), meaning that no full occurrence has yet been seen. 
More formally, let $i$ be the current position, and let us first assume that $\real(v)=1$. Then 
$\minmax(v) = x$ if $x = \min \{\max C[s\dd s+\sd(v)-1] : T[s\dd s+\sd(v)-1] = L(v) \text{ and } s+\sd(v)-1<i\}$, and $\txtpos(v) = s_0$ for one such $s_0$, i.e., (i) $T[s_0\dd s_0+\sd(v)-1]=L(v)$,  (ii) $s_0+\sd(v)-1<i$, and (iii) $\max C[s_0\dd s_0+\sd(v)-1] = x$. 

Now let us look at the case $\real(v)=0$, we have yet to see an occurrence of $L(v)$. Initially,  $\minmax(v)=+\infty$; when we encounter a non-empty prefix of $L(v)$, of length $0<d\leq\sd(v)$, starting, say, in position $s_0$, we update $\minmax(v)$ to $\max C[s_0\dd s_0+d-1]$ and $\txtpos(v)$ to $s_0$. Thus, we have seen an occurrence of a prefix of $L(v)$ but not yet a full occurrence of $L(v)$, and we are optimistic since we are hoping to find a full occurrence whose max does not exceed the current one. However, as soon as we find the first full occurrence (and set $\real(v)=1$), from that point on we only update $\minmax(v)$ and $\txtpos(v)$ if we see another full occurrence. Therefore, $\real(v)$ is updated exactly once during the algorithm.

\subparagraph*{Finding an admissible phrase and choosing its source.} 
Let us now assume that we have processed $T[1\dd i-1]$ and want to find the next phrase and source. We match $T[i\dd]$ in the suffix tree, making sure during navigation that we only get admissible prefixes of $T[i\dd]$. In particular, if we are in node $v$ and should go to child $u$ of $v$ next (because $T[i+\sd(v)]$ is the first character of the edge label $(v,u)$), then we first check if $\minmax(u) < c$. If so, then we can descend to $u$ and continue from there, skipping over the next $\sd(u)-\sd(v)$ positions in $T$. Otherwise, $\minmax(u)\geq c$ and we 
return the new phrase $(\txtpos(v),\sd(v),T[i+\sd(v)])$. Moreover, the $C$-array for $j=i, \ldots, i+\ell$ is set according to Def.~\ref{def:chainlength}.

\subparagraph*{Updating the suffix tree annotation.} After the new phrase has been computed, we need to update the annotations in the suffix tree. For $j\leq i + \ell$, going backward in the string, we will update the nodes on the leaf-to-root path from leaf $j$. The idea is the following. 

Fix $j \leq i+\ell$. The prefix $T[j\dd i+\ell]$ of $T[j\dd]$ now has the $C$-array filled in, so its max-value $m = \max C[j\dd i+\ell]$ is known. This may or may not necessitate updates in the nodes on the path from leaf $\leaf_j$ to the root. First, for the leaf $j$ itself, if $i\leq j$, then the $\minmax$ is still $+\infty$, so we set $\minmax(\leaf_j)\gets m$. Otherwise, we are seeing a longer prefix of $T[j\dd]$ than before, so we update $\minmax(\leaf_j)\gets \max(\minmax(\leaf_j),m)$. Regarding the nodes $v$ on the path from $\leaf_j$ to the root: their labels are increasingly shorter prefixes of suffix $T[j\dd]$, so they need to be updated only as long as $j+\sd(v)-1\geq i$, since otherwise, the prefix $L(v)$ does not overlap with the newly assigned subinterval $C[i\dd i+\ell]$. 

So let $j+\sd(v)\geq i$, there are two cases. First, if $j+\sd(v)-1\leq i+\ell$, then $m$ is a realistic value, since the entire corresponding $C$-array interval has been filled in. Therefore, we can then compute $m$ in a more clever way by using an RMQ on $C$, i.e., $m = RMQ(C, j, j+\sd(v)-1)$. So if $\real(v)=0$, then we update $\minmax(v) \gets m$ and $\real(v) \gets 1$. Otherwise (if $\real(v)=1$), an update is needed only if $\minmax(v) > m$, in which case we set $\minmax(v) \gets m$ and $\txtpos(v) \gets j$; since $\real(v)=1$, we have seen the label $L(v)$ before and already had a realistic value for its $\minmax$ value. Second, if $j+\sd(v)-1 > i+\ell$, then $m$ is an optimistic value only, and therefore, we update the annotation of $v$ only if $\real(v)=0$; in that case, we set $\minmax(v) \gets m$ and $\txtpos(v)\gets j$. 

Finally, we use the following criterion for how far back in the string we need to go with $j$. 
If no node in the path from $\leaf_j$ to the root can be effected by the new phrase, then we do not need to consider position $j$ at all in the current iteration. This holds if the label of the parent node does not reach $i$, i.e., if $j+\sd(\parent(\leaf_j))-1 < i$. We compute an auxiliary array $E[1\dd n]$ s.t.\ $E[j] = j+\sd(\parent(\leaf_j))-1$. 
It is easy to see that $E[j]\leq E[j']$ if $j<j'$.
This means that, moving back-to-front, we can stop at the first $j$ for which $E[j]<i$.

A worst-case time complexity for a Minmax parse producing $z'$ phrases is $O(z' n^2) \subseteq O(n^3)$, as in principle one can consider every $j \in [1\dd i-1]$ for every new phrase $T[i\dd i+\ell]$, traverse the $O(n)$ ancestors of $\leaf_j$, and run an RMQ operation on each. While the RMQ structure we use on $C$ is dynamic, it only undergoes appends to the right, in which case it is possible to support updates in $O(1)$ amortized time and queries in $O(1)$ time \cite[p.~5]{Fis11}. We do not know if this cubic complexity is tight, however. In practice we expect $z'$ to be much less than $n$ on highly repetitive texts, and the height of the suffix tree to be logarithmic, yielding a time complexity of $O(z'n\log n)$, which thus becomes practical on repetitive data.

\begin{example}\label{ex:greedier}
In Fig.~\ref{fig:st-greedier}, we can see the suffix tree $\ST$ for $T={\tt alabaralalabarda\$}$ with some additional annotations in some nodes. In this example, the first three phrases, i.e., ${\tt a}$, ${\tt l}$, and ${\tt ab}$, have been already computed, with the corresponding chain lengths in $C$ and annotations in $\ST$. The annotations exhibit non-trivial updates using the colour red, namely updates that are different than changing the starting value for $\minmax$, i.e., changing $\minmax$ from  $+\infty$ to a finite value. Nodes whose annotation is not shown have not been updated yet, therefore, they have $\minmax=+\infty, \txtpos=-1,$ and $\real=0$. The updates caused by the new phrase ${\tt ar}$ are highlighted in blue. First, to find the longest previous factor we have to descend to the child with label ${\tt a}$, then we check whether the child with label ${\tt ar}$ has $\minmax < c$. In this case, it was not less than $c$ ($\minmax = +\infty$), so we stopped the search and output the new phrase ${\tt ar}$. Then all suffixes $j$ with $1 \leq j \leq 6$ undergo an update starting from the corresponding leaf; e.g., leaves $5$ and $6$ and corresponding ancestors get updated to some non-initial value, whereas inner nodes with label ${\tt abar}$, ${\tt alabar}$, ${\tt bar}$ and ${\tt labar}$ change $\real$ to $1$ because $j + \sd(v) - 1 \leq i+\ell$.
\end{example}

\input{figure_st}

\section{The Greedier Parser: Combining Greedy with Minmax}\label{sec:greedier}

We now combine the ideas of the Greedy and the Minmax parsers, using the enhanced suffix tree to consider only longest admissible phrases. Consider when the Minmax algorithm stops in a node $v$ and returns $(\txtpos(v),\sd(v),T[i+\sd(v)])$. It did not descend to the next child $u$ because $\minmax(u)=c$, i.e., every occurrence of $L(v)$ seen so far has a value $c$ somewhere in the $C$-array. However, it is possible that in one of these occurrences, the position of this $c$ is after $L(v)$; in other words, that we could have gone down the edge some way towards $u$. 

To check this, we will use the $D$-array from Sec.~\ref{sec:greedy}, in addition to the $C$-array and the enhanced suffix tree. Let $v$ and $u$ be as before, i.e., $v$ is parent of $u$, $\minmax(v) < c$, $\minmax(u)\geq c$, $L(v)$ is a prefix of $T[j\dd]$ and $T[j+ \sd(v)]$ is the first character of the label of $(v,u)$. Let $d$ be the maximum value of $D[k]$ for some occurrence of $L(u)$ that we have already processed, so $d$ is the largest distance from the start of an occurrence of $L(u)$ to the next $c$ in the $C$-array. We return $(\txtpos(v),\sd(v),T[i+\sd(v)])$, as before, if $d\leq \sd(v)$, and $(k, D[k], T[i+D[k])$, where $k$ is a leaf in $u$'s subtree with $D[k]=d$, otherwise. As for updating the annotations, if node $v$ has $\minmax(v) = c$ and some $\txtpos(v) = x$, then, when performing the traversal from $\leaf_j$ up to the root, we want to change $\txtpos(v) \gets j$ if $D[j] > D[\txtpos(v)]$.
It is easy to see that the Greedier algorithm now returns the longest admissible phrases; otherwise, it works similarly to the Minmax algorithm. The time complexity increases
to $O(z'n^2\log n) \subseteq O(n^3\log n)$, because we need dynamic RMQs on array $D$ as well, which undergoes updates at arbitrary positions.


\section{Experiments}\label{sec:experiments}


We implemented the \BLZ\ parsing algorithms in C, and ran our experiments on an
AMD EPYC 7343, with 32 cores at 1.5 GHz, with a 32 MB cache and 1 TB of RAM. 
We used the repetitive files from Pizza\&Chili ({\tt
http://pizzachili.dcc.uchile.cl}) and compared the number of phrases produced
by \BLZ\ using different maximum values $c$ for the chains,
with the number of phrases produced by LZ (i.e., with no limit $c$). We used a classic LZ implementation \cite{KKP13cpm} where the source of each phrase is its lexicographically closest suffix.\footnote{It is likely that using the variant called ``rightmost LZ parse'', which chooses the rightmost source, gives better results because it tends to distribute the uses of the sources more uniformly. Such a parse seems to be nontrivial to compute \cite{BP16,EFP23}, however, and we are not aware of practical implementations.}

As a reference point, we also implemented two simple baselines that obtain a
\BLZ\ parse. The first, called $\BLZ$1, runs the classic LZ parse and then cuts the 
phrases at the points where the chain lengths reach $c+1$. Since the symbol becomes explicit, its chain length becomes zero and the chain lengths of the symbols referencing it decrease by $c+1$. We should then find the new positions that reach $c+1$, and so on. It is not hard to see that this laborious postprocessing can be simulated by just adding, to the original $z$ value of LZ,  
the number of positions $i$ where $C[i] \bmod (c+1) = 0$. 

The second baseline,
$\BLZ$2, is slightly stronger: when it detects that it has produced a text
position exceeding the maximum $c$, it cuts the phrase there (making the
symbol explicit), and restarts the LZ parse from the next position. This gives
the chance of choosing a better phrase starting after the cut, unlike
$\BLZ$1, which maintains the original source.

Despite some optimizations, our Greedy \BLZ\ parser consistently reaches the
$\Theta(\log^3 n)$ time complexity per text symbol, making it run at about 3 MB per minute. The Greedier and the Minmax parsers, despite their cubic worst-case time complexity, run at a similar pace: 1.9--4.7 MB per minute: our upper bound is utterly pessimistic, and perhaps not tight. 

Table~\ref{tab:colls} shows the main characteristics of the 
collections chosen. We included two versioned software repositories ({\tt coreutils}
and {\tt kernel}, where the versioning has a tree structure), two versioned
documents ({\tt einstein} and {\tt leaders}, where the versioning has a
linear structure), and two biological sequence collections ({\tt para} and {\tt
influenza}, where all the sequences are pairwise similar). The average phrase
length is in the range 160--365 and the maximum chain length of a
symbol is in the range 38--70. The exception is {\tt einstein}, which is
extremely compressible and also has a very large $c$ value. 

As a point of comparison, the table also includes the grammar size and height obtained with a balanced version of RePair \cite{LM00}.\footnote{From {\tt www.dcc.uchile.cl/gnavarro/software/repair.tgz}, directory {\tt bal/}.} We modified the RePair grammar so as to remove the nonterminals that are referenced only once, inserting their right-hand side in that unique referencing place. The maximum grammar height is comparable with $c$ as a measure of access cost in the grammar-compressed text. We can see that the height is considerably smaller than $c$, for the price of a weaker compression method.

\begin{table}[t]
\small
\begin{center}
\begin{tabular}{l@{~}|@{~}r@{~}|@{~}r@{~}|@{~}r@{~}|@{~}r@{~}|@{~}r@{~}|@{~}r@{~}|@{~}r}
File & \multicolumn{1}{c@{~}|@{~}}{$\sigma$}	& \multicolumn{1}{c@{~}|@{~}}{$n$}	    &
\multicolumn{1}{c@{~}|@{~}}{$z$}
& \multicolumn{1}{c@{~}|@{~}}{$n/z$}    & max $c$ & \multicolumn{1}{c@{~}|@{~}}{$g$}  & $h$ \\
\hline
{\tt coreutils} & 236 & $205{,}281{,}779$ & $1{,}286{,}070$ & $160$ & $66$	& $2{,}409{,}429$ & $28$ \\ 
{\tt kernel}	& 160 & $257{,}961{,}616$ & $705{,}791$     & $365$ & $70$ & $1{,}374{,}651$ & $32$ \\ 
{\tt einstein}  & 139 & $467{,}626{,}545$ & $75{,}779$      & $6{,}171$ & $1{,}736$ & $212{,}902$ & $47$ \\ 
{\tt leaders} 	& 89 & $46{,}968{,}181$ & $155{,}937$  & $301$ & $60$ & $399{,}667$ & $27$ \\ 
{\tt para}      & 5 & $429{,}265{,}758$ & $1{,}879{,}635$ & $228$ & $38$ & $5{,}344{,}477$ & $26$ \\ 
{\tt influenza} & 15 & $154{,}808{,}555$ & $557{,}349$     & $278$ & $63$ & $1{,}957{,}370$ & $26$ \\ 
\hline
\end{tabular}
\end{center}
\caption{Our repetitive text collections and some statistics:
alphabet size $\sigma$, length $n$, number $z$ of phrases in the LZ parse, average phrase length
$n/z$, maximum chain length in our LZ parse, size $g$ of a balanced grammar, and height $h$ of that grammar.
\label{tab:colls}}
\end{table}

\begin{figure}[t!]
\centering
\includegraphics[width=0.49\textwidth]{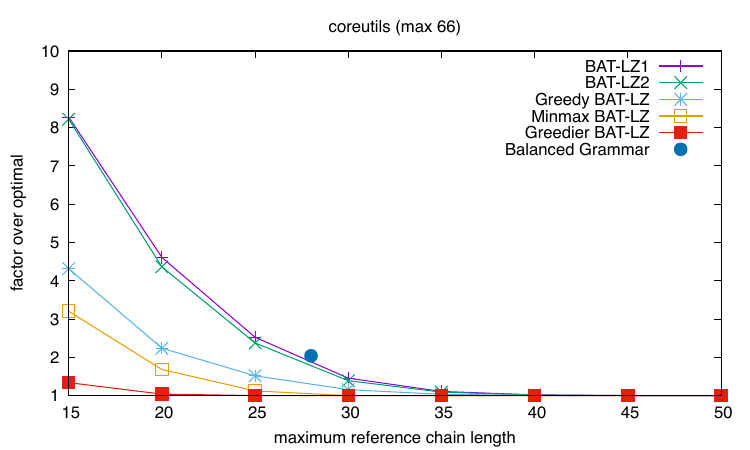}
\includegraphics[width=0.49\textwidth]{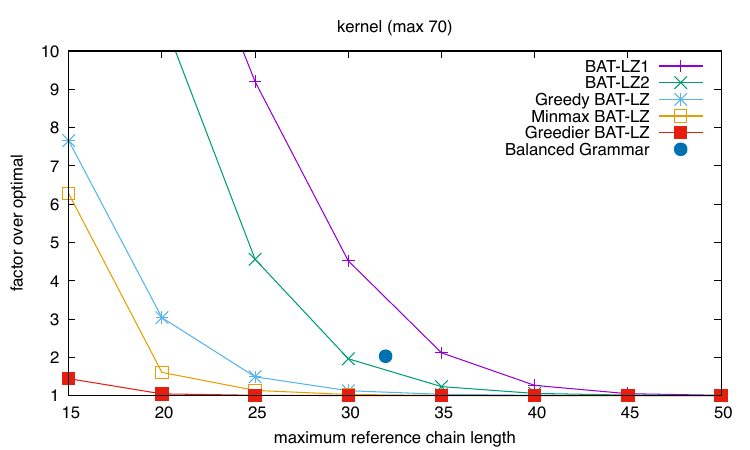}

\includegraphics[width=0.49\textwidth]{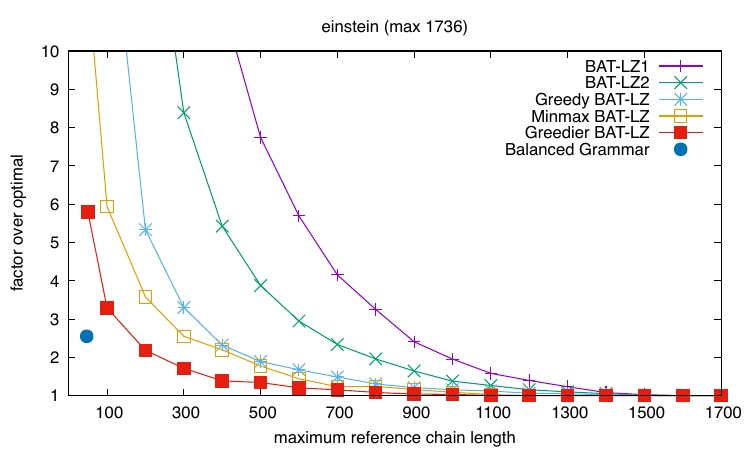}
\includegraphics[width=0.49\textwidth]{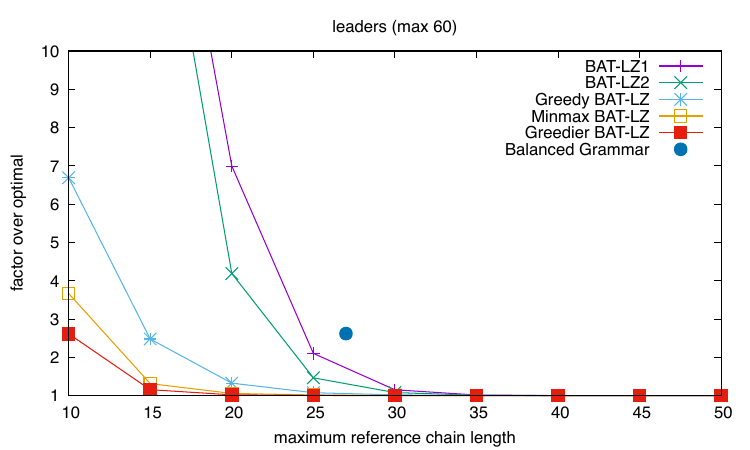}

\includegraphics[width=0.49\textwidth]{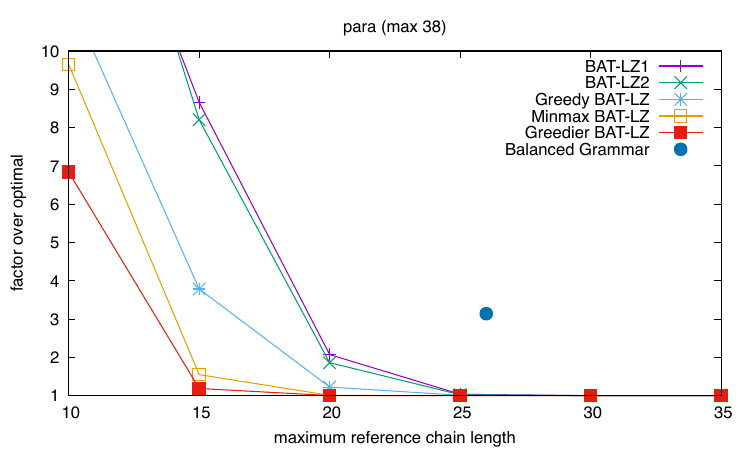}
\includegraphics[width=0.49\textwidth]{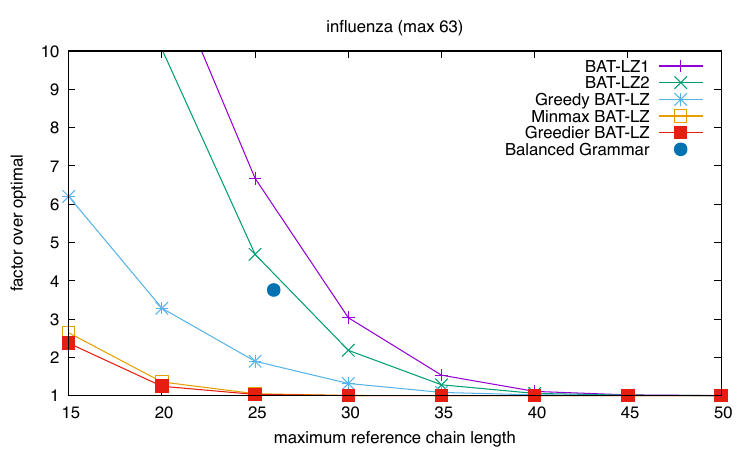}

\caption{Overhead factor of number of \BLZ\ versus LZ phrases as a function of
the maximum length $c$ of a chain, for our different \BLZ\ parsers and a balanced grammar.}
\label{fig:plot}
\end{figure}

Fig.~\ref{fig:plot} shows how the quotient between the number of phrases
generated by the \BLZ\ parsers and by the optimal number of LZ phrases 
evolves as we allow longer chains. It can be seen that our Greedy
\BLZ\ parser sharply outperforms the baselines in terms of compression
performance. Our Greedy parser is, in turn, outperformed by Minmax, and Minmax is outperformed by our Greedier parser. The last one 
reaches a number of phrases that is only 1\% over the
optimal for 
$c$ as low as 20--30, which is 0.7--1.1 times $\log_2 n$. 

We also show in the figures the balanced grammar method, using the values of Table~\ref{tab:colls}.\footnote{For a fair comparison of space, we consider a tight space needed to support fast extraction: For each of the $z$ phrases we count $\log_2 n$ bits to point to the source, $\log_2(n/z)$ bits for the length (as there are $z$ lengths adding up to $n$; the cumulative sequence of lengths also allow finding the desired phrase using Elias-Fano codes \cite{Eli74,Fan71}), and 8 bits for the final symbol. For a grammar of size $g$ and $r$ symbols, we count $g\log_2 r$ bits for the right-hand sides, $g\log_2 n$ bits for the expansion lengths (cumulative on the right-hand sides to binary search them), and $r\log(g/r)$ bits to encode the rule lengths with Elias-Fano.} We can see that grammars are competitive, in some cases, with the simple baselines, but not with our new algorithms, which yield much better tradeoffs. 
The only exception to this analysis is {\tt einstein}, which features a huge maximum $c$ 
value of $1{,}736$ and whose (extremely low) $z$ value is approached only
with $c$ values near 700 using our \BLZ\ parsers. On this text, the balanced grammar offers an access time that is not achievable with our techniques.

\begin{figure}[t!]
\centering
\includegraphics[width=0.49\textwidth]{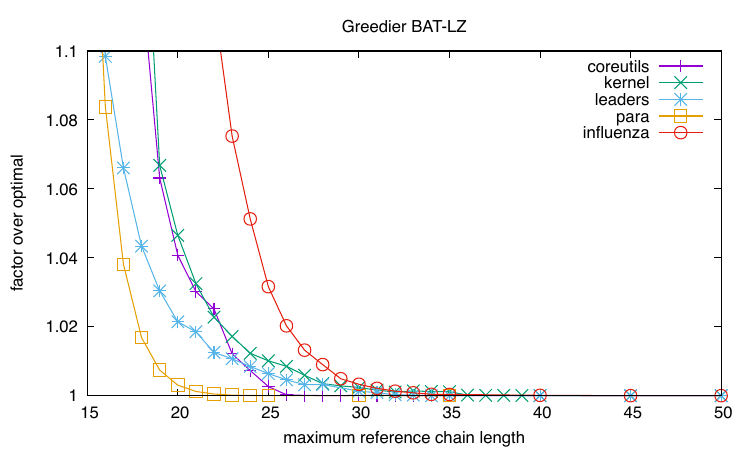}
\includegraphics[width=0.49\textwidth]{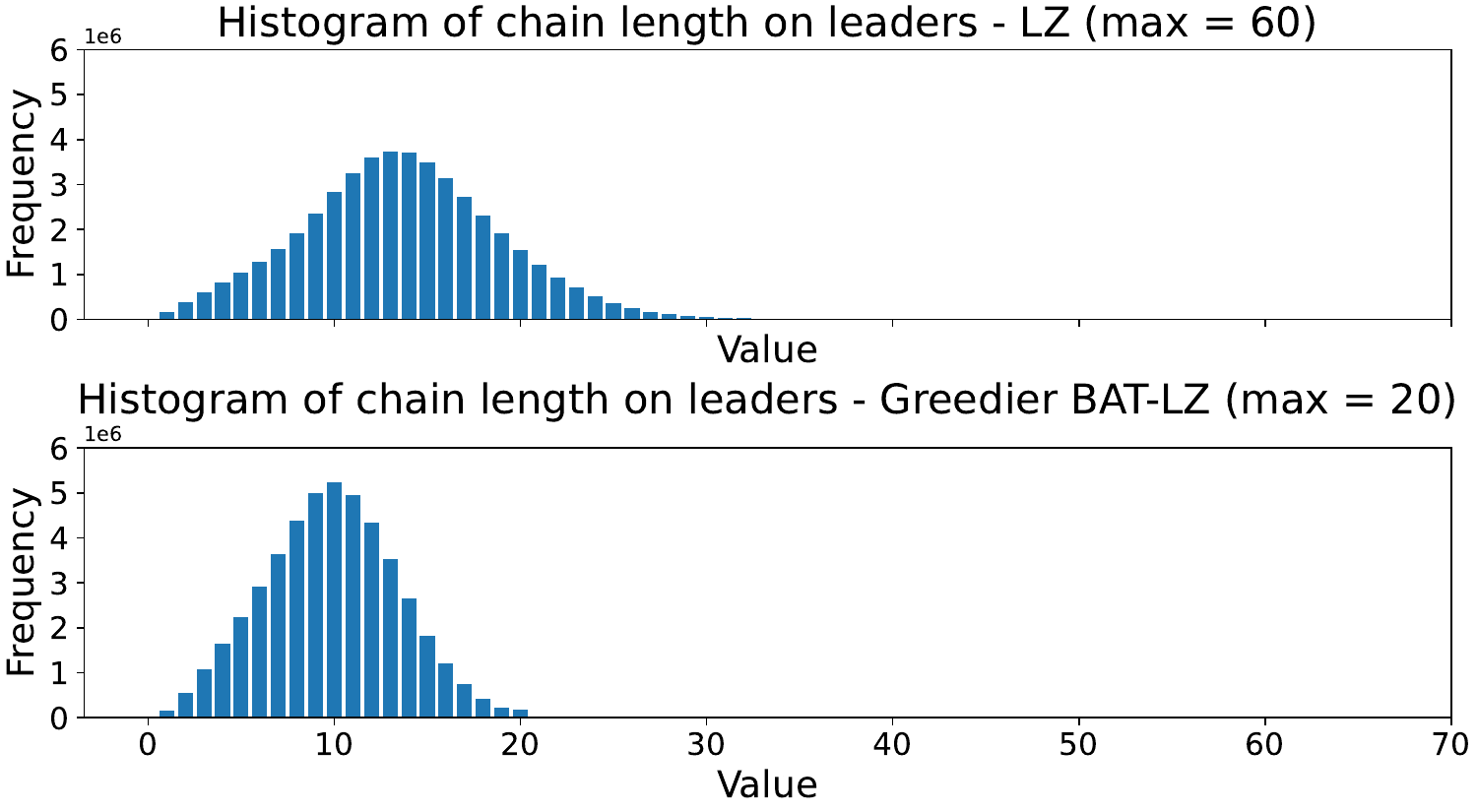}

\caption{Left: detail of Fig.~\ref{fig:plot}, for the Greedier \BLZ\ parser,  
focusing on the overheads below 10\% over the LZ parse. Right: a comparison of histograms with shared $x$ and $y$ axis representing the chain length values on {\tt leaders}; LZ on top and Greedier \BLZ\ with $c = 20$ on the bottom.}
\label{fig:greedier}
\end{figure}

Fig.~\ref{fig:greedier} (left) zooms in the area where Greedier \BLZ\   
reaches less than 10\% extra space on top of standard LZ (excluding {\tt einstein}).

\section{Discussion and Future Work}

A first question is whether a Greedy \BLZ\ parsing can be produced in $o(n\log^3 n)$ time within linear space, either by solving our geometric problem faster or without recasting the parse into a geometric problem. This question seems to be answered in a very recent work, simultaneous with ours, that gives an $O(n\log\sigma)$-time 
greedy algorithm \cite{BannaiFHMP} based on simulating a suffix tree construction.\footnote{They use a slightly modified definition of Lempel-Ziv parses, which has no explicit character at the end of the phrases. The precise consequences of this difference are not totally clear to us.} This algorithm is likely to be faster than ours in practice, but also to use much more space, which is relevant when compressing large repetitive texts. They 
also propose a parse similar to our BAT-LZ2, along with others that are incomparable to ours (in particular to Greedier, our best performing BAT-LZ parse).

Besides our reduction to a geometric problem being of independent interest, we believe that its flexibility can be exploited to compute more sophisticated parses in $O(n\log^3 n)$ time. For example, it might compute the Greedier parse if we extend the RMQ data structure to incorporate the additional optimization criterion (minmax of sources). 

Other heuristics may also be of interest: there may be better ways to rank sources, other than their maximum chain length. 
Further, we have so far focused on reducing the worst case access time, but we might prefer to reduce the {\em average} access time. Our parsings do reduce it (Fig.~\ref{fig:greedier} right), but this is just a side effect and has not been our main aim. So we pose as an open problem to efficiently build a leftward parse with bounded average reference chain length whose number of phrases is minimal, or in practice close to that of classical LZ. 

Another intriguing line of work is to study the compression performance of BAT-LZ. An important result by Bannai et al.~\cite{BannaiFHMP} shows that, letting $g_{rl}$ be the size of the smallest run-length context-free grammar that generates a text $T$, there exists a BAT-LZ parse for $T$ of size $O(g_{rl})$ if we let $c=\Theta(\log n)$ with some convenient multiplying constant. This bound is nearly optimal, because existing bounds \cite{VY13} forbid the existence of BAT-LZ parses of size $O(g)$---where $g \ge g_{rl}$ is the size of the smallest context-free grammar---with access time $c=O(\log^{1-\epsilon}n)$ for a constant $\epsilon>0$. A relevant question is whether there is a BAT-LZ parse of size $O(z)$---where $z \le g_{rl}$ is the size of the Lempel-Ziv parse of $T$---with $c=\Theta(\log n)$.

Finally, from an application viewpoint, it would be interesting to
incorporate \BLZ\ in the construction of the LZ-index~\cite{KN13} and
measure how much its time performance improves at the price of an
insignificant increase in space. Obtaining an efficient bounded version of the
LZ-End parsing described in the same article~\cite{KN13} is also an
interesting problem since\shorten{, as mentioned,} efficient parsings for unrestricted 
LZ-End have appeared only recently~\cite{KK17,KK17b}.

\newpage

\bibliography{paper}

\end{document}